\documentclass[UTF8]{IEEEtran}
\usepackage{lineno}
\usepackage{amssymb}
\usepackage{amsfonts}
\usepackage{amsmath}
\usepackage{algorithm}
\usepackage{algorithmic}
\usepackage{tikz}
\usepackage{graphicx,subfigure,cite,multicol,multirow,diagbox,booktabs,array}
\usepackage{color}
\usepackage{bm}
\usetikzlibrary{shapes.geometric, arrows}
\ifCLASSINFOpdf
\else
\fi

\begin{document}
\title{Physical Layer Security Techniques \\for Future Wireless Networks}
\author{Weiping Shi, Xinyi Jiang, Jinsong Hu, Yin Teng, Yang Wang,\\
Hangjia He, Rongen Dong, Feng Shu, Jiangzhou Wang}
\maketitle
\begin{abstract}
The broadcast nature of wireless communication systems makes wireless transmission extremely susceptible to eavesdropping and even malicious interference. Physical layer security technology can effectively protect the private information sent by the transmitter from being listened to by illegal eavesdroppers, thus ensuring the privacy and security of communication between the transmitter and legitimate users.  The development of mobile communication presents new challenges to physical layer security research. This paper provides a comprehensive survey of the physical layer security research on various promising mobile technologies, including directional modulation (DM), spatial modulation (SM), covert communication, intelligent reflecting surface (IRS)-aided communication, and so on. Finally, future trends and the unresolved technical challenges are summarized in physical layer security for mobile communications.
\end{abstract}


\IEEEpeerreviewmaketitle

\section{Introduction}
Wireless mobile communications has been developing very fast \cite{wang1,wang2,wang3,wang4,wang5}. The fifth generation (5G) mobile systems have been started to be deployed worldwide. Due to the broadcast nature of wireless media, information security has been a critical issue in wireless communication. The traditional method of addressing communication security is to adopt the secret key encryption. The secure transmission of private data is achieved by designing various encryption algorithms in the upper protocol stack. However, cryptography is of computational security and its security level depends on the hardness of the underlying mathematical problem it employs. Once an effective method is found to solve its mathematical problem, the security of the encryption method will be seriously compromised \cite{wuyongpeng}.

\begin{figure}
  \centering
  \includegraphics[width=0.3\textwidth]{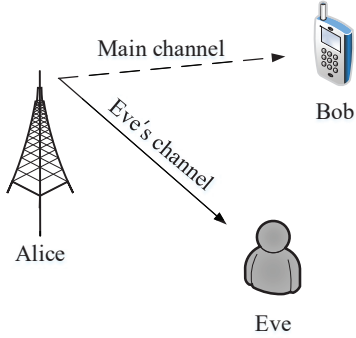}
  \caption{A three-node secure communication model.}
  \label{three-node_Mod}
\end{figure}

Physical layer security technology has become an effective solution to the wireless communication security problem. As shown in Figure .\ref{three-node_Mod}, the transmitter (i.e., Alice) sends a confidential message to the legitimate receiver (i.e., Bob), while the eavesdropper (i.e., Eve) receives the signal and intends
to decode it. The key idea of physical layer security technology is to exploit the inherent propagation characteristics of wireless channels (such as the difference between the main channel and the eavesdropping channel, randomness, and reciprocity) from the perspective of information theory.
By reasonably designing the transmit signal so that it improves the amount of mutual information between the transmitter and the desired user at the physical layer while reducing the amount of information in the eavesdropping channel. Therefore, the physical layer security technology can effectively protect the content of private messages sent by the transmitter from being eavesdropped by illegal eavesdroppers \cite{chenxiaoming}. This improves the privacy and security of the transmitter's communication with other legitimate users. Compared with traditional encryption technology, physical layer security technology has the following notable features and advantages: First, the physical layer security technology can achieve keyless security, that is, no encryption and decryption operations are required. Second, physical layer security techniques can take advantage of the time-varying and random nature of wireless channels. Signal processing techniques are employed to design reasonable beamforming or power allocation strategies from the transmitter's perspective to improve the security performance of wireless communication systems \cite{wangdong}. Based on the above advantages, solving the communication security problem from the physical layer has aroused widespread concern.
The authors in \cite{Leun_27} studied the secure transmission of private information on Gaussian channels, and proved that expanding the difference between the main channel and the eavesdropping channel can achieve low probability interception and low probability detection for eavesdroppers.
\cite{wang_tifs28} further researched the keyless physical layer secure transmission technology over fading channels.
The use of multiple antennas can add additional degrees of freedom and further improve the security performance of the wireless network \cite{Ng_29,Jeo_30}. In addition, by generating random artificial noise (AN) at the transmitter to interfere with eavesdroppers, the security of the system can be further improved  \cite{Goel_31}.
Currently, scenarios where eavesdroppers exist are considered in various wireless communication systems.

\begin{figure}
  \centering
  \includegraphics[width=0.45\textwidth]{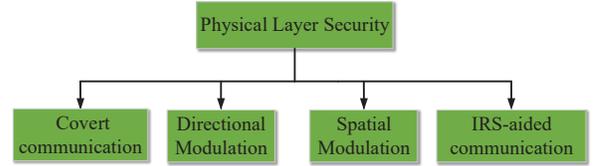}
  \caption{An illustration of physical layer security.}
  \label{Four_phy}
\end{figure}

Considering the potential of physical layer security for mobile communications, the opportunities and challenges of how to achieve high levels of security at the physical layer deserve more attention from the research community. The purpose of this paper is to provide a comprehensive summary of the latest physical layer security research results for key future wireless network technologies. As shown in Figure \ref{Four_phy}, we will focus on the following four aspects of physical layer security technologies.

\begin{enumerate}
  \item Covert communication: In covert communication, when the transmitter transmits a message to the receiver, it is guaranteed that the probability that the illegal watcher can detect the transmission is small enough. Covert transmission technique, as an important secure transmission technique, aims to hide the transmission behavior of the transmitter. Compared to physical layer security techniques that aim to prevent the transmitted content from being overheard by eavesdroppers, covert communication techniques can achieve a higher level of communication security.
  \item Directional modulation (DM): The directional modulation technique using phased arrays sends information along the direction of desired receiver, making the constellation diagram of the received signal in the desired receiver direction the same as that of the baseband modulated signal, while the constellation diagram of the received signal in the undesired direction will be distorted, thus ensuring the safe transmission of information.
  \item Spatial modulation (SM): Spatial modulation, a new multiple-input multiple-output (MIMO) antenna transmission technique, plays an effective balance between hardware overhead and transmission rate. Because it makes full use of the channel index information to transmit more bit streams, it enhances the spectral efficiency of the system without increasing the costly  radio frequency (RF) links. Since both antenna index and modulation symbols in SM networks carry private information, interception of either of them may result in leakage of confidential information. In other words, compared with MIMO communication, SM network has serious communication security risks. Therefore, it is strategically important to utilize physical layer security technology to provide security for SM systems.
  \item Intelligent reflective surface (IRS)-aided communication: An IRS is an artificial surface made of electromagnetic material and composed of a large number of passive reflective units. By configuring these reflective units to act on the phase shift and amplitude of the incident signal, fine-grained three-dimensional beamforming can be achieved, which can be used to improve channel quality, enhance received power, and extend the communication distance. IRS transforms the traditional uncontrollable and random wireless communication environment into a programmable and relatively deterministic transmission space, and plays an active role in the signal transmission process. By introducing IRS into the security system, the security of the system can be further enhanced.
\end{enumerate}

The rest of this paper can be outlined as follows. We first discuss covert communication in Section II. Then, the cases of secure DM systems are researched in Section III. In Section IV, we study SM technique on physical layer security. Section V focuses on the security in IRS-aided wireless communication systems. The future research directions and conclusions are presented in Section VI and Section VII respectively.

\section{Covert communication}
While most information-theoretic security works to date have revolved around the issues of protecting the content of information, the growing concern around mass communication surveillance programs has reignited interest in investigating the covertness of communications, also known as low probability of detection (LPD), in which covertness requires that the communication remains undetectable from a watchful adversary, named warden. The concept of covert communication has shown tremendous research value and application value \cite{Bash2015hiding,Shihao2019LPD}. In the covert communication scenarios, the transmitter (Alice) possesses sensitive information that needs to be sent to the legitimate receiver (Bob), while the warden (Willie) listens to the communication environment and tries to detect any covert transmission from Alice to Bob with some probable existed uncertainties. These uncertainties can be formed from Willie's receiver noise power, imperfect channel knowledge, finite blocklength, and interference from other network users or artificial noise.

Willie faces a decision as to whether or not Alice sent any covert information to Bob, which means that Willie faces a binary hypothesis testing problem.
The null hypothesis states that Alice did not transmit while the alternative hypothesis states that Alice did transmit, sending covert information to Bob. We define the probability of false alarm (or Type I error) as the probability that Willie makes a decision in favor of the alternative hypothesis, while the null hypothesis is true. Similarly, the probability of miss detection (or Type II error) is define as the probability of Willie making a decision in favor of the null hypothesis, while the alternative hypothesis is true. The detection error probability at Willie is the sum of Type I error and Type II error.
In the following, we present a review of literatures considering the covertness constraint in term of detection error probability or Kullback-Leibler (KL) divergence for achieving covert communication, where KL divergence is a statistical distance of how one probability distribution of null hypothesis is different from another probability distribution of the alternative hypothesis.

\subsection{Covert Communication with Relay}
The ultimate goal of covert communication is to achieve shadow wireless networks. Preliminary results towards this goal have been achieved in the context of relay networks. Covert communication with the aid of relay  was analyzed in \cite{Hu2018covertrelay,Jinsong2019Selfsustained,Chan2021Covert,Sheikholeslami2018multihop}. The scenario investigated in \cite{Hu2018covertrelay} is that there is a source node which attempts to transmit information to a destination node via a relay node.  The relay node receives a message from a source node and then attempts to retransmit the source's original message in addition to its own covert message to the destination without the source node detecting the covert transmission. This work presents two schemes (rate-control and power-control) for evaluating the probability that the covert criteria is satisfied.

In \cite{Jinsong2019Selfsustained}, the authors studied the covert communications with a self-sustained relay over Rayleigh fading channels, where the relay harvests extra energy from the source to transmit covert message to destination by employing a more powerful energy harvester with a higher conversion efficiency factor, while the source is to detect the covert communication. The work aims to find the maximum throughout, given a certain level of covertness for two energy harvesting schemes, time switching and power splitting. Covert communication with a positive rate was shown to be possible in the transmission from a relay node to a destination when the source is uncertain regarding the forwarding strategy of the relay node.

Two relay selection schemes (i.e., random selection and superior-link selection) in multiple relays-assisted internet of things (IoT) systems was proposed in \cite{Chan2021Covert}, where the source needs to transmit two types of messages with/without covert requirement. The numerical results showed that the superior-link selection scheme has significant improvement in term of covert capacity when compared to the random selection under the same transmission power at the source. The routing of information in covert wireless networks with multiple relay nodes and multiple wardens under additive white gaussian noise (AWGN)channels has been considered in \cite{Sheikholeslami2018multihop}, where the algorithms were developed to find the path with the maximum throughput and the path with minimum end-to-end delay when considering a single key and the case of independent keys at the intermediate nodes.

\subsection{Covert Communication with Full-duplex}
In the covert communication, there is always a trade-off between the throughput and covertness. Using a full-duplex receiver offers a two-fold benefit for the covert communication relative to generating AN by a separate and independent jammer. Firstly, it enables a higher degree of control over the transmitted AN signals (e.g., its power), thus causing further deliberate confusion at the warden. Secondly, the cutting-edge self-interference cancellation techniques can be adopted to provide higher data rates for covert communications with the full-duplex receiver.
In \cite{Khurram2018fullduplex}, the transmitter Alice aims to send message at a fixed rate to a legitimate receiver Bob over a fading wireless channel while the whole communication is under the surveillance of the warden Willie. Operating at the full-duplex mode, Bob can receive information from Alice and also simultaneously inject interference to confuse Willie over the same channel, which can increase the detection error probability at Willie. On the other hand, it can also decrease covert throughput due to the effect of self-interference at Bob. Based on these discussions, the trade-off between the achievable rate, the transmission power, and covertness criterion that is the probability of error minimized over all prior distributions of the fading channel was investigated.

The aforementioned works in the literature mainly focus on how to hide the wireless transmission action rather than the transmitter itself, since there is some prior information that can indicate the existence of the transmitter. To achieve higher level covert communication, the transmitter itself also should be hidden from the warden. Taking advantage of the full-duplex and channel inversion power control (CIPC), the work aims to hide the transmitter by removing the requirement on channel state information (CSI) \cite{hu2018CIPC}. For the CIPC schemes adopted in this work,  the transmitter varies its transmitted signal power as per the channel from itself to the receiver, so that the received power of the covert information signal is a fixed value. Therefore, neither Bob nor Willie knows the corresponding CSI perfectly, since Alice does not transmit any pilot signal at all. Meanwhile, the full-duplex receiver can further confuse the detection of Willie.

Ultra-reliable and low-latency communications (URLLC) is envisioned to support mission critical applications with stringent requirements of latency and reliability, where the codeword is required to be short, e.g., the order of 100 channel uses. Specifically, with the aided of the full-duplex receiver, the authors in \cite{Feng2019Delay} showed that transmitting AN with a fixed power is still helpful in covert communication with finite blocklength, since in a limited time period the warden cannot exactly learn its received power. The blocklength has significant impact on  the detection performance at Willie and the maximal achievable rate of the channel from Alice to Bob (i.e., the maximal achievable rate decreases as blocklength  decreases for a fixed decoding error probability).

\subsection{Covert Communication with UAV}
For the static covert communication network, the positions of the three participants (Alice, Bob and Willie) are fixed. The unmanned aerial vehicle (UAV) assisted communication networks enable the participants to realize more flexible deployments in covert communication \cite{Jiang2021UAV}. UAV can establish line-of-sight (LoS) wireless links for air-ground communications, which provides significant performance improvement over the conventional non-line-of-sight (NLoS) terrestrial communications. In some scenarios, the covert rate can be effective improved by exploiting the mobility of UAV while the detection by Willie will become easier due to the component of LoS channel.

In \cite{Xiaobo2019covertUAV}, a UAV acted as a mobile relay to enhance covert transmission by dynamically adjusting its trajectory and transmit power. Specifically, the authors formulated an optimization problem that maximizes the average covert transmission rate subject to the transmission outage probability constraint at Bob and covertness constraint at Willie as well as the UAV's mobility constraint and transmit power constraint. Such a joint optimization problem is generally difficult to tackle directly due to the non-convex constraints. To solve this optimization problem, the authors first transformed the intractable transmission outage probability constraint into a deterministic form by using the conservative approximation and then applied the first-order restrictive approximation to transform the optimization problem into a convex form, which is mathematically tractable.
A multi-hop relaying strategy (e.g., the number of hops, transmit power) was optimized to maximize the throughput against the surveillance of a UAV warden in \cite{Wang2020UAV}.
The beam sweeping based detection of a UAV's transmission by a terrestrial warden with multiple antennas was investigated in \cite{hu2020UAV}, where the Pinsker's inequality and Kullback-Leibler divergence were adopted to derive the detection error probability.

\begin{figure} [ht!]
    \begin{center}
    \includegraphics[width=0.4\textwidth]{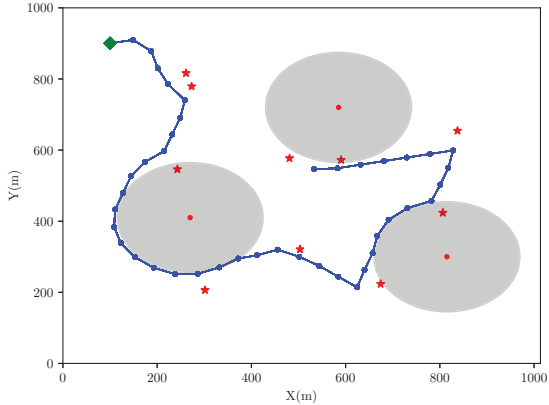}
    \caption{The UAV's optimized trajectory.}\label{fig5}
    \end{center}
\end{figure}

\begin{figure} [ht!]
    \begin{center}
    \includegraphics[width=0.4\textwidth]{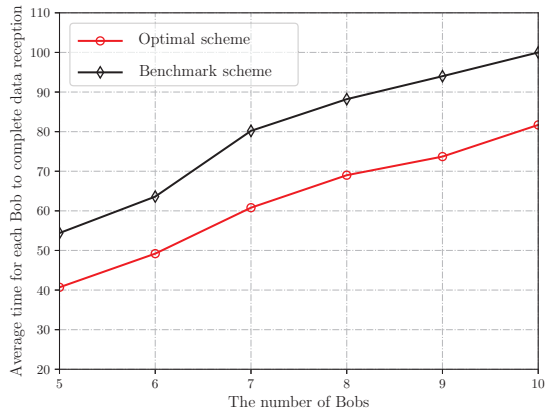}
    \caption{Average time for each Bob to complete data reception achieved by our TD3-CDD scheme and the benchmark scheme.}\label{fig6}
    \end{center}
\end{figure}

In the aforementioned works, the UAV's trajectory designs with the optimization for some parameters may suffer from several limitations, for example, the objective function or constraints in the optimization problems are non-convex and difficult to be tackled. Against this background, the deep reinforcement learning (DRL) has been serving as an efficient solution to handle the decision-making issue due to its essential traits, i.e., learning dynamically from the real world. Motivated by the superiorities of DRL, we intend to address the non-convex problem for considering covert data dissemination with UAV, where UAV intends to transmit the covert data to some legitimate receivers (Bobs) under the surveillance by some wardens (Willies). Specially, a twin-delayed deep deterministic policy gradient-based covert data dissemination (TD3-CDD) algorithm is proposed to minimize the time of covert data dissemination by jointly optimizing the UAV's trajectory and Bobs' schedule. In details, we first determined the covertness constraint explicitly by analyzing the detection performance at each Willie. Then, we formulated an optimization problem to minimize the total time for data dissemination subject to the covertness constraint and other practical constraints, e.g., the UAV's mobility constraints. By applying the deep reinforcement learning approach, an efficient algorithm named TD3-CDD is developed to solve the optimization problem.

In Figure.~\ref{fig5}, we plot the optimized UAV's trajectory achieved by our proposed TD3-CDD scheme. where the locations of the Bobs are indicated by ``$\star$", the green diamond indicates the initial position of the UAV, the central locations of Willies are indicated by ``$\bullet$" and the shaded area indicates the surveillance area of each Willie. In this figure, the UAV is able to sequentially visit all the Bobs while avoiding each Willie's surveillance. Figure.~\ref{fig6} shows the average time for each Bob to complete the data reception achieved by our TD3-CDD scheme and the benchmark scheme (e.g., UAV's flight trajectory of keeping away from Willies without optimization). In this figure, as expected, we observe that the TD3-CDD scheme takes less time than the benchmark scheme to complete the data transmission to all Bobs.

\section{Physical layer security of DM}
  In recent years, as one of the key technologies of physical layer wireless transmission, DM has been a research hotspot in the field of physical layer security (PLS). DM technology realizes the directionality of signal transmission by optimizing the design of beamforming and adding AN to the transmitted signal. Using this technology can not only accurately and efficiently transmit information to desired users, but also achieve signal constellation distortion in undesired directions by designing AN projection matrix to transmit AN to undesired directions. Different from traditional physical layer security technologies, directional modulation is mainly suitable for fading-free Gauss white noise channels. When the linear distance and direction of the transmitter and receiver are known, the channel steering vector has static stability.

\begin{figure}[htp]
\centering
\includegraphics[width=0.4\textwidth]{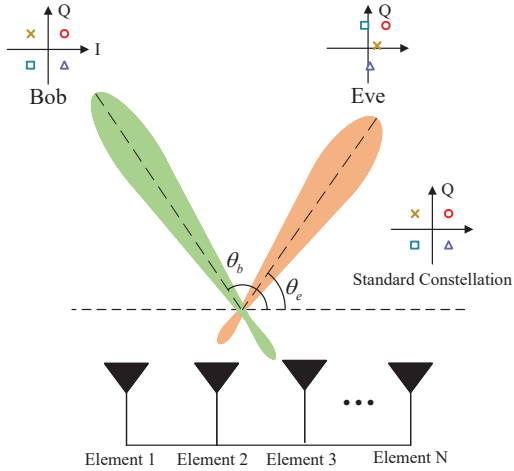}
\caption{Schematic diagram of DM network.}
\label{Sys_Mod}
\end{figure}

  In Figure.~\ref{Sys_Mod}, a schematic diagram is given to show the basic idea of DM. The system consists of three nodes: a multi-antenna transmitter (Alice), a desired user (Bob) and an eavesdropper (Eve). By using the AN technology and the null space projection (NSP) criterion, the AN is projected into the null space of the desired direction channel, that is, it does not affect the desired user, so that the expected receiver can easily decode useful information. At the same time, since AN makes it difficult for the eavesdropper to observe the changing law of the signal constellation diagram, the eavesdropper cannot correctly decode confidential messages (CMs).

\subsection{Realization method of DM}
DM can be divided into two categories: one is based on the combination of radio frequency end components, and the other is implemented by designing algorithms for baseband signals. {Babakhani} \emph{et al.} \cite{Babakhani} introduced a near-field direct antenna modulation (NFDAM) technology, which modulates the far-field radiation signal through the time-varying electromagnetic boundary conditions of the antenna's near-field. This enables the transmitter to send information depending on the direction, so as to achieve a safe transmission effect. {Daly} and {Bernhard} \cite{Daly2009} proposed a DM technique based on phased array generation. By correctly phase shifting each element, the amplitude and phase required by each symbol in the digital modulation scheme are generated in a given direction. The disadvantage of this technology is that it only considers the signal constellation in the desired direction and ignores the consideration of the distortion of the constellation in the undesired direction. Therefore, {Daly} \emph{et al.} \cite{Daly2010} designed a phased array-based physical layer secure transmission technology on the basis of \cite{Daly2009}. They found through experiments that the DM transmitter creates a narrower low error rate area around the desired direction, while continuing to maintain a high error rate in the sidelobe area. {Daly} and {Bernhard} \cite{Daly201058} then used an array with pattern reconfigurable elements to switch elements for each symbol, thereby generating a digitally modulated signal in the desired direction, and used this reconfigurable array to demonstrate the improvement of DM in terms of safety performance. Unlike traditional transmitters that excite the same antenna, {Hong} \emph{et al.} \cite{Hong2011} proposed a dual-beam DM technology for PLS communications, which uses in-phase and quadrature baseband signals to excite two different antennas.

\subsection{AN-aided DM Secure Systems}
According to the concept of orthogonal AN in information theory, {Ding} \emph{et al.} applied orthogonal vector technology to DM in \cite{Ding2014}. By adding orthogonal AN signals at the baseband, the signal constellation in the undesired direction is distorted, causing eavesdroppers to be unable to decipher CMs according to the changing law of the signal constellation. At the same time, the orthogonal AN is projected into the null space (NS) of the channel steering vector of the desired direction, thereby eliminating the influence of AN on the signal received by desired users. {Ding} \emph{et al.} \cite{Ding2015} further proposed an orthogonal vector method for DM transmitter to synthesize multiple beams on the basis of  \cite{Ding2014}. \cite{Ding2016} showed that when the non-cooperative receivers are placed separately, the DM system can be regarded as a MIMO system operating in free space. By transmitting AN which orthogonal to useful information, the freedom of spatial multiplexing is lost, thereby enhancing the security performance of the DM system.

\subsection{Robust beamforming and DOA measurement for DM Secure System}
In actual application scenarios, there will be errors in the measured angle, which will affect the expected user reception performance and reduce the security of wireless transmission. However, in the above research, the direction angle is assumed to be ideal known information, thus further study on systems with imperfect information of direction angle is needed. {Hu} \emph{et al.}  \cite{Hu2016} designed a robust DM synthesis technique based on the principle of minimum mean square error (MMSE) in view of the uncertainty of the estimated direction angle, which improves the bit error rate (BER) performance of the system by minimizing the constellation distortion in the desired direction. {Shu} \emph{et al.} \cite{Shu2016} considered a multi-beam broadcast DM scenario with an imperfect desired direction. \cite{Shu2016} designed beamforming by minimizing information leakage in the eavesdropping direction, and maximizing the expected average received signal-to-interference-to-noise ratio (SINR) of the user end to design a projection matrix of AN, so as to achieve the robustness of CMs transmission in the wireless communication system. {Shu} \emph{et al.} \cite{Shu2018Robust} considerd the MIMO system with direction of arrival (DOA) measurement error, and proposed a robust beamforming scheme based on the combination of main lobe integration and leakage, which transmits independent and parallel private data streams to multiple legitimate users at the same time, thereby realizing the stable and safe transmission of CMs in the DM system.  {Shu} \emph{et al.}  further designed a low-complexity scheme  in \cite{Shu2018Secure} that can accurately and safely transmit CMs to desired users through the joint use of AN, phase alignment, orthogonal frequency division multiplexing (OFDM) technology, and DM random subcarrier selection technology.

In addition, there is a key technology in the DM field, DOA estimation technology, which combines information of different frequencies to obtain the best DOA. {Hung} and {Kaveh} \cite{Hung1988} characterized the specific structure of the focus matrix to avoid focus loss, and the focus matrix is preferably a unitary matrix, which has an important impact on the statistical characteristics of DOA estimation. In \cite{Claudio2001}, the data was robustly preprocessed through the weighted average of the signal subspace and the enhanced design of the focus matrix. {Ng} \emph{et al.} \cite{Ng2005} proposed a wideband structure for processing array signals, and estimated the order and DOA through the Bayesian method and the reversible jumping Markov chain Monte Carlo process.
\section{Physical layer security of SM}
As one of the derivatives in the context of MIMO techniques, SM is becoming a hot research area. Its basic idea is using both the transmit antenna index and the modulated symbol to convey messages, which improves the spectrum efficiency and reduces the cost of radio frequency chain \cite{RY2008}. Compared to MIMO technique, SM plays a crucial role in balancing the transmit rate and the hardware cost. On the other hand, the broadcast nature of the wireless communication incurs that the desired receiver is vulnerable to hostile users, thus establishing a set of systematically secure transmission strategies becomes an imperative demand. In this section, we review the crucial transmission strategies of the physical layer security under the typical SM system and hybrid SM system respectively.
\subsection{Typical SM system}
\begin{figure*}[!t]
\centering
\includegraphics[width=0.9\textwidth]{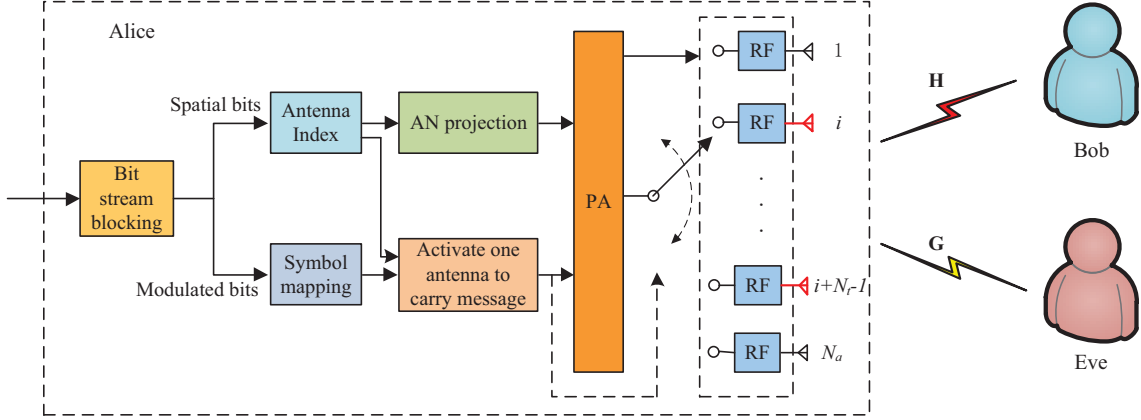}
\caption{Schematic diagram of SSM network.}
\label{SMsystem}
\end{figure*}

A typical SM system has been shown in Figure.~\ref{SMsystem}. In this figure, four important tools including transmit antennas selection (TAS), AN projection, power allocation (PA) and  receive beamformer at the desired receiver are fully employed to achieve a high-performance secure SM (SSM).

In Figure.~\ref{SMsystem}, the transmitter (Alice) is equipped with $N_a$ transmit antennas (TAs). Without loss of generality, $N_a$ is not a power of two, thus $N_t$ out of $N_a$ TAs should be selected for mapping binary bits to the antenna index. Appropriately selecting out an active antenna group is capable of improving the security performance of SM systems. The authors in \cite{EDAS2013} proposed two TAS methods: capacity-optimized antenna selection (COAS) and Euclidean distance-optimized antenna selection (EDAS) which focused on improving bit-error rate performance. In \cite{wang2018}, the authors generalized EDAS method to the secure SM system and proposed the maximizing signal-to-leakage-plus-noise ratio (Max-SLNR) TAS strategy, where the Max-SLNR scheme achieved a higher secrecy rate (SR) performance with a lower complexity. However, Max-SLNR method dose not directly maximize the secrecy rate of the SSM system. To overcome the above disadvantage, the authors in \cite{xia2019} proposed two low-complexity TAS methods for maximizing SR by analyzing the asymptotic performance of SR when signal-to-noise ratio (SNR) approaches 0 and $\infty$. Furthermore, a compromise solution that suits for the whole SNR region was further devised, which has the capability of achieving a close SR performance when compared with the exhaustive search (ES) method. When a rough CSI of Eve's channel is obtained, the authors in \cite{xiatasan} proposed a simulated-annealing-mechanism-based TAS scheme which obtained excellent SR profit.

By reasonably generating Gaussian AN at the transmitter, the detection of the Eve can be deteriorated, so as to further improve the security performance of the system \cite{AN}. \cite{NSP} projected AN onto the null-space of the desired channel to interfere unknown eavesdropper. Although NSP scheme ensures the high security performance in high SNR region, the beamformer design for AN has not been further expored to further improve the security performance of SSM. The authors in \cite{xiatasan} proposed a novel ratio optimization algorithm which reduced computational overhead and achieved near-optimal safety performance.

As the total power is constrained, elaborately splitting the power allocated to the CM and allocated to the AN has a great impact on improving the SR performance. In \cite{liuPA} and \cite{xiaPA}, the approximation of SR based on cut-off rate was invoked to substitute the non-closed SR expression for secure SM systems. Using the approximation as the objection function, the authors in \cite{liuPA} proposed a gradient-based PA method which approached the optimal PA method based on ES. In \cite{xiaPA}, a novel criterion that maximizes the ratio of Bob's SLNR and of Eve's AN-to-leakage plus noise ratio (ANLNR) was conceived which was close to the optimal PA factor and also reduced some computational complexity. To further reduce the computational complexity and approach the optimal SR performance, the authors in \cite{shu2021spatial} proposed a deep-neural-network (DNN)-based PA strategy whose key idea was to treat the input and output of ES algorithm as an unknown nonlinear mapping and used a DNN to approximate it.

Consider a SSM system with a full-duplex (FD) malicious attacker (Mallory), where Mallory works on FD mode, eavesdropping from Alice and transmitting jamming to Bob simultaneously. A game-theoretic approach was adopted in \cite{choi2016full} to deal with a combination of passive eavesdropping and active attacks. The authors in \cite{jiang2021efficient} proposed three effective receive beamformers with low complexity to eliminate jamming and improve SR.

\subsection{Hybrid SM system}
When the number of TAs tends to large-scale, the circuit cost and complexity will become a burden on fully-digital SSM. To address this problem, hybrid SM was proposed and fully investigated where the total antenna array is divided into multiple transmit antenna subarrays with each being connected to single RF chain. As for hybrid SSM, two effective tools including precoder and transmit antenna subarray selection are used to achieve a high security SSM.

\begin{figure}[htbp]
\centering
\includegraphics[width=0.5\textwidth]{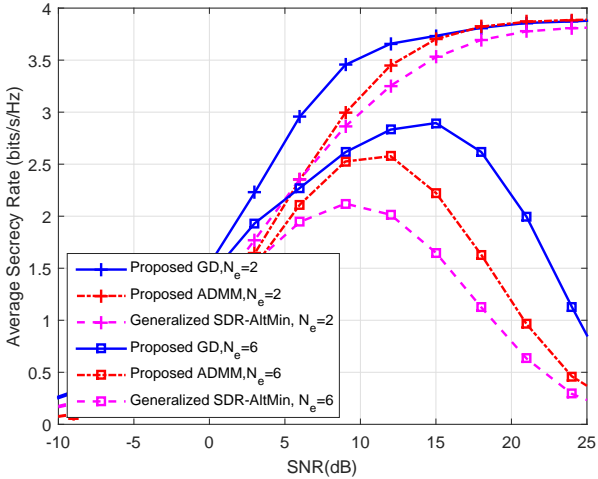}
\caption{Comparison of SR performance of various precoders.}\label{Comparison-of-SR-No-AN}
\end{figure}

In \cite{yu2016alternating}, the authors proposed a semi-definite relaxation based alternating minimization (SDR-AltMin) algorithm to jointly design the digital and analog precoders. In \cite{xiaoyu2021}, the authors generalized SDR-AltMin algorithm to the hybrid SSM system, and proposed two precoders which maximized approximate SR (ASR) based on gradient method and alternating direction method of multipliers respectively. Figure.~\ref{Comparison-of-SR-No-AN} shows that the proposed two methods in \cite{xiaoyu2021} harvested higher SR performance than the generalized SDR-AltMin method. For hybrid generalized SSM systems, the authors in \cite{xiaprecoder} proposed a novel algorithm to separately optimize the analog precoder and the digital precoder to maximize SR. As the objective function is non-concave, the authors conceived a pair of concave objective functions to optimize the analog precoder and digital precoder.

As the number of RF chains is not a power of two, there exists the transmit antenna subarrays selection problem. The authors in \cite{xiaoyu2021} proposed two low-complexity transmit antenna subarray selection (TASS) methods, named maximizing eigenvalue (Max-EV) and maximizing product of signal-to-interference-plus-noise ratio and artificial noise-to-signal-plus-noise ratio (Max-P-SINR-ANSNR), which performed better than the generalized Max-SLNR method.

\section{IRS-aided physical layer security}

\begin{figure}[htb]
  \centering
  \includegraphics[width=0.4\textwidth]{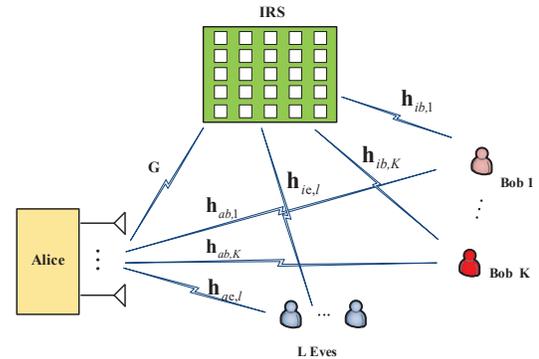}
  \caption{An IRS-aided secure multi-user communication system.}
  \label{IRS_Sys_Mod}
\end{figure}

Recently, IRS has been considered as a promising new technology for the next-generation wireless communications, which can reconfigure the wireless propagation environment via controlling reflection with software \cite{wu2019towards}. Specifically, an IRS composes a large number of low-cost passive reflecting elements, each of which can induce a phase change to the incident signal independently. By smartly adjusting the phase shift of the reflecting elements, the IRS reflected signals and the signals from other paths can be combined constructively to enhance the desired signal power or destructively to suppress undesired signals such as co-channel interference, which thus significantly improves the communication performance. A typical IRS-aided MISO multi-user secure system is shown in Figure. \ref{IRS_Sys_Mod}. The introduction of IRS in the system can increase the achievable rate of legitimate users (i.e., Bobs) while suppressing the achievable rate of illegal users (i.e., Eves), ultimately improving the security performance of the system. %The introduction of IRS can boost the secrecy data rate (DR) by increasing the DR of the legitimate receiver while decreasing the DR of the eavesdropper (EVE).
Hence, IRS can be used for strengthening the system security under the wiretap channel, especially when the channel of the eavesdropping communication link is stronger than that of the legitimate link.

\begin{figure}
  \centering
  \includegraphics[width=0.4\textwidth]{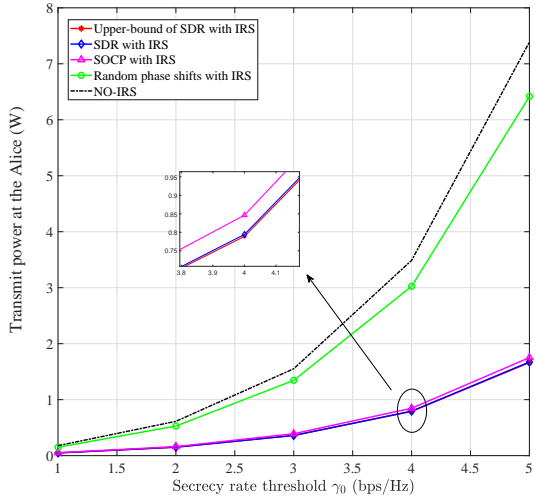}
  \caption{Transmit power at the Alice versus SR.}
  \label{IRS_P_SR}
\end{figure}

\subsection{IRS-aided MISO Secure Systems}
Many recent studies have utilized IRS to secure the physical layer of wireless communications.
The authors in \cite{cui2019secure,shen2019secrecy,2019Enabling} investigated an IRS-aided secure wireless system where a multi-antenna transmitter communicates with a single-antenna receiver in the presence of an Eve.
In \cite{cui2019secure,shen2019secrecy,2019Enabling}, the SR was maximized by jointly optimizing the transmit beamforming and the IRS phase shifts. Specifically, the authors in \cite{cui2019secure} proposed an alternating optimization (AO) algorithm to design two variables alternately, in which the optimal solution to the transmit beamforming was computed directly and the IRS phase shifts were optimized by using semidefinite relaxation (SDR) method. In \cite{shen2019secrecy}, the transmit beamforming and the IRS phase shifts were also optimized in an alternating manner. In each iteration, the solution to the transmit beamforming was achieved in closed form, while a semi-closed form solution to the IRS phase shifts was obtained by adopting the Majorization-Minimization (MM) algorithm. The element-wise block coordinate descent (BCD) and AO-MM algorithms were developed for solving the problem efficiently in \cite{2019Enabling}. %The two algorithms were more suitable for the wireless systems with small-scale IRSs and large-scale IRSs, respectively. Different from the above three papers, the design goal in \cite{chu2019intelligent}, \cite{feng2019secure} was to minimize the transmit power under the SR constraint. In particular, the closed-form solution of the optimal transmit beamforming was derived and the rank-relaxation method for the phase shift design was presented in \cite{chu2019intelligent}. The authors in \cite{feng2019secure} considered two scenarios, including rank-one and full-rank channels of the access point (AP)-IRS links. On the one hand, a low-complexity design was facilitated conveniently due to the advantage of rank-one channel. On the other hand, the eigenvalue-based algorithm was combined with the SDR/projected gradient descend algorithm in the case of full-rank channels.
%Qiao \emph{et al}. \cite{qiao2020secure} investigated the SR performance of an IRS-aided millimeter-wave (mmWave) and terahertz (THz) system. Under the discrete phase-shift and rank-one channel assumption, the close-form solution of the transmit beamforming was directly derived, and the optimal phase shifts were obtained by utilizing the SDR-based method and element-wise BCD-based method.
Furthermore, the authors in \cite{chen2019intelligent} investigated a minimum-SR maximization problem in a secure IRS-aided multiuser multiple-input single-output (MISO) broadcast system with multiple Eves. The problem was successfully solved by applying the path-following algorithm and AO technique in an iterative manner. Moreover, two suboptimal algorithms with closed-form solutions were developed to further reduce the computational complexity.
To further enhance the security, the AN is designed to disturb Eve.
In \cite{guan2020intelligent}, AN was firstly considered in an IRS-aided secure communication system. Specifically, the achievable SR of the system was maximized by jointly optimizing the transmit beamforming with AN and IRS phase shifts. An efficient algorithm based on AO was developed to solve the problem sub-optimally. The authors in \cite{xu2019resource} considered the resource allocation design in an IRS-aided multiuser MISO communication system. Aiming to maximize the system sum SR, the beamforming vectors, AN covariance matrix at the BS and phase shift matrix at the IRS were jointly optimized by applying AO, SDR and manifold optimization. The authors in \cite{shi2021secure} investigated a secure IRS-aided multigroup multicast MISO communication system to minimize the transmit power subject to the SR constraints. First, an SDR-based AO algorithm was proposed and a high-quality solution was obtained. Then, to reduce the high computation complexity, a low-complexity AO algorithm based on second-order cone
programming (SOCP) was presented. Simulation result indicates that these designs can guarantee reliable
secure communication with the aid of IRS as shown in Figure. \ref{IRS_P_SR}.
\subsection{IRS-aided MIMO Secure Systems}
Recently, the introduction of IRS in MIMO system to enhance the security  has also made effective progress.
Dong \emph{et al}. \cite{dong2020secure} and Jiang \emph{et al}. \cite{jiang2020intelligent} studied an IRS-aided secure  MIMO communication system, where a base station (BS) equipped with multiple antennas communicates with a multi-antenna receiver in the presence of a multi-antenna Eve. Particularly, the joint design of the transmit covariance at the BS and the phase shift coefficients at the IRS was considered to maximize the SR. In \cite{dong2020secure}, the barrier, Newton and backtracking line search methods were used to search for global optimal transmit covariance, while the MM algorithm was applied to obtain the local optimal phase shift coefficients. In \cite{jiang2020intelligent}, the successive convex approximation (SCA)-based algorithm was used to solve the optimization of the transmit covariance. For the design of the phase shift coefficients, the closed-form solution of each phase shift coefficient was individually provided in an alternating way while fixing other phase shift coefficients. Hong \emph{et al}. \cite{hong2020artificial} considered an AN-aided secure MIMO communication system assisted by an IRS. With the aim for maximizing the SR, the BCD algorithm was exploited to alternately optimize the variables. Specifically, the optimal transmit covariance and AN covariance matrix were derived in semi-closed form by applying the Lagrangian multiplier method, and the optimal phase shifts were obtained in closed form by utilizing the MM algorithm. IRS-aided secure DM-MIMO system was studied in \cite{DM_IRS_MIMO}. Two confidential bit streams transmitted from Alice to Bob, different from the conventional DM where Alice only transmits single CBS to Bob. This paper demonstrated that the IRS significantly enhanced the SR of DM.

%Liu \emph{et al}. \cite{liu2020energy} and Niu \emph{et al}. \cite{hehao2020intelligent} studied a secure IRS aided simultaneous wireless information and power transfer (SWIPT) system, where the information receiver (IR) and energy receiver (ER) are all equipped with multiple antennas, and the ER is regarded as a potential EVE. The design goal in \cite{liu2020energy} was to maximize the energy efficiency subject to the transmit power constraint, the unit modulus constraint and the harvested energy constraint. SDR was adopted to tackle the non-convexity of the optimization problem and then AO was employed to obtain the sub-optimal solution. To address the SR maximization problem in \cite{hehao2020intelligent}, an inexact BCD based algorithm was proposed, in which the unit modulus and harvested energy constraints were solved by applying the penalty MM and complex circle manifold methods.
\subsection{Robust IRS-aided Secure System}
The above contributions are obtained under the assumption that perfect CSI is available at the Alice. However, obtaining perfect CSI is ideal, so it is necessary to consider robust security performance of the system.
The authors in \cite{lu2020robust,yu2020robust,hong2020robust} investigated the robust transmission design in an IRS-aided secrecy system, where the CSI of Eves' channels is not perfectly known. Particularly, a robust secure beamforming problem was formulated to maximize the worst case of SR in \cite{lu2020robust}. By replacing the wiretap channels with a weighted combination of discrete samples and then using SDR technique, an efficient AO algorithm was proposed to solve the optimization problem sub-optimally. In \cite{yu2020robust}, the authors considered a scenario where two IRSs are deployed to improve the sum SR of multiple single-antenna legitimate receivers in the presence of multiple multi-antenna Eves. Considering the imperfection of the CSI of the Eves' channels, a robust non-convex optimization problem was proposed. Then the design of the transmit covariance, AN covariance matrix and IRS phase shifts was handled by adopting AO, SCA and SDR. In \cite{hong2020robust}, the authors first considered the statistical CSI error model and the imperfect cascaded channels of AP-IRS-EVE in IRS-aided secure communications. Then an outage-constrained power minimization problem was formulated. To solve it, the Bernstein-type inequality was exploited to tackle the outage rate probability constraints, and then a sub-optimal algorithm based on AO, the penalty-based and SDR methods were utilized to optimize the variables alternately.

\section{Future Technical Challenges and Research Directions}
Despite many researches in studying the physical layer security, there are still many challenging problems to be solved. In the following,  some technical challenges for physical layer security in wireless communications are discussed.

\subsection{CSI Acquisition of IRS-aided communication system}

The acquisition of perfect channel information in physical layer security is a challenging task. It is mainly manifested in the following aspects. First, it is especially difficult to obtain the CSI associated with the Eve due to the hidden nature of eavesdroppers. Second, in IRS-assisted security systems, because of the limited signal processing capability at the IRS, the perfect CSI of IRS-links is challenging to obtain. Hence, designing a secure and reliable CSI feedback mechanism is necessary.

\subsection{DOA Acquisition of DM}
DM technology has been applied to key safety tools in vehicle communications, IoT, UAVs, smart transportation and deep-space communications.
Due to the gathering effect of AN, DM is only applicable to LoS channels, not to multipath channels. Therefore, how to extend DM to mmWave MIMO channels with multipath is still a challenging problem in the future. Besides, since there may be a large angle error when sending CMs to the desired user at a long distance, it is more practical to study the robust DM synthesis technology in the future for the uncertainty of the direction angle estimation.

\subsection{Channel Model of SM Aided by IRS}
The previous research on SM and IRS aided SM system are based on the Rayleigh fading model. How to extend SM system to other types of fading channels is very important for the practical application and deployment of SM. In addition, how to construct the security optimization problem in the relevant channel environment is worth pondering.

\section{Conclusion}
The emergence and development of future wireless technologies such DM/SM technology, covert communication, and IRS have brought new security challenges to future networks. Designing effective and secure transmission schemes for future wireless communications that exploit the propagation properties of radio channels in the physical layer has recently attracted extensive research interest. This paper reviewed the currently popular physical layer secure communication techniques from both theoretical and technical perspectives. We investigated the confidentiality issues of physical layer security technologies such as DM, SM, covert communication, and IRS, respectively. Finally, we discussed potential challenges in physical layer security and point out some possible future research directions.

%%%%%%%%%%%%%%%%%%%%%%%%%%%%%%%%%%%%%%%%%%%%%%%%%%%%%%%
%%% Acknowledgements. ÖÂл
%%%%%%%%%%%%%%%%%%%%%%%%%%%%%%%%%%%%%%%%%%%%%%%%%%%%%%%
%\Acknowledgements{This work was supported in part by the National Natural Science Foundation of China (Nos. 62071234, and 62071289), the Hainan Major Projects (ZDKJ2021022),  the Scientific Research Fund Project of Hainan University under Grant KYQD(ZR)-21008 and KYQD(ZR)-21007, and the National Key R\&DProgram of China under Grant 2018YFB1801102.}

\ifCLASSOPTIONcaptionsoff
  \newpage
\fi
\bibliographystyle{IEEEtran}
\bibliography{IEEEfull,cite}
\end{document}